\documentclass[a4paper,12pt]{article}
\usepackage{epsf,amsmath}
\usepackage[dvips,usenames]{color}
\usepackage{graphicx}
\usepackage{color}
\usepackage{colortbl}

\newlength{\dinwidth}
\newlength{\dinmargin}
\newcommand{\spur}[1]{\not\! #1 \,}

\newcommand{\half}{\frac{1}{2}}

\begin{document}

\title{\bf  Constraints on the anomalous tensor operators
from $B\to\phi K^{\ast}, \eta K^{\ast}$ and $\eta K$ decays}

\author{Qin Chang$^{ab}$, Xin-Qiang Li$^{cd}$,
Ya-Dong Yang$^a $\\
{$^{a}$ \small Institute of Particle Physics, Huazhong Normal University, Wuhan, 
Hubei  430079, P. R. China}
\\
{ $^b$\small Department of Physics, Henan Normal University,
Xinxiang, Henan 453007, P.~R. China}
\\
{ $^c$\small Institute of Theoretical Physics, Chinese Academy of
Sciences, Beijing, 100080, P.~R. China}
\\
{ $^d$\small Graduate School of the Chinese Academy of Sciences,
Beijing, 100039, P.~R. China}
\\
 {\small E-mail: changqin1981@163.com,  xqli@itp.ac.cn,  yangyd@iopp.ccnu.edu.cn} \\
 }

\maketitle

\bigskip\bigskip

\maketitle \vspace{-1.5cm}

\begin{abstract}
We investigate whether the anomalous tensor operators with the
Lorentz structure $\sigma_{\mu\nu} (1+\gamma_{5})\otimes
\sigma^{\mu\nu}(1+\gamma_{5})$, which could provide a simple
resolution to the polarization anomaly observed in $B\rightarrow
\phi K^{\ast}$ decays, could also provide a coherent resolution to
the large ${\cal B}(B\to \eta K^{\ast})$ and survive bounds from
$B\to\eta K$ decays. Parameter spaces satisfying all these
experimental data are obtained, and found to be dominated by the
color-octet tensor operator contribution. Constraints for the
equivalent solution with $(1+\gamma_{5})\otimes(1+\gamma_{5})$
operators are also derived and found to be dominated by the
color-singlet one. With the constrained parameter spaces, we finally
give predictions for $B_{s}\to \phi \phi$ decay, which could be
tested at the Fermilab Tevatron and the LHC-b experiments.

\end{abstract}


\newpage

\section{Introduction}

Looking for signals of physics beyond the Standard Model~(SM) is one
of the most important missions of high energy physics. Complementary
to direct searches for new physics~(NP) particles in the high energy
colliders, the study of $B$ physics is of great importance for
probing indirect signals of NP. In this respect, the $B$ factories
at SLAC and KEK are doing a commendable job by providing us with a
huge amount of data on various $B$-meson decays,  which are mostly
in perfect agreement with the SM predictions. However,
there still exist some unexplained puzzles, such as the unmatched $CP$ asymmetries in
$B\rightarrow \pi K$ decays~\cite{PDG06,HFAG,th_pik}, the abnormally
large branching ratios of $B\rightarrow \eta^{\prime}K$ and
$B\rightarrow \eta K^{\ast}$
decays~\cite{PDG06,HFAG,Richichi:1999kj,babar2,belle2,th_etak,Yang:2000xn,
beneke1,kundu1}, and the large transverse polarization fractions in
$B\rightarrow \phi K^{\ast}$
decays~\cite{PDG06,HFAG,babar1,belle1,cdf1,th_phik,Beneke:2006hg,kagan,KC,YY,kundu2}.
Confronted with these anomalies, we are forced not only to consider
more precise QCD effects, but also to speculate on the existence of
possible NP scenarios beyond the SM.

It is well-known that the flavor-changing neutral current~(FCNC)
processes arise only from loop effects within the SM, and are
therefore very sensitive to various NP effects. Since the puzzling
$B$-meson decay processes mentioned above are all related to the
FCNC $b\to s$ transitions, these decay channels could be used as
effective probes of possible NP scenarios. So, if one kind of NP
could resolve one of these puzzles, it is necessary to investigate
whether the same scenario can also provide a simultaneous resolution
to the others. Considering the fact that current theoretical
estimations of ${\cal B}(B\to\eta^{\prime} K)$ still suffer from
large uncertainties~\cite{beneke1,flavorsinglet}, and the NP
scenario with anomalous tensor operators considered in this
paper do not contribute to the $B\to \phi K$ and $B\to \pi K$ decays
in the naive factorization~(NF) approximation, we shall only focus
on the $B\to \eta K^{(\ast)}$ and $B\to \phi K^{\ast}$ decays.

The recent experimental data on the longitudinal polarization
fraction $f_{L}$ in $B^{0}\rightarrow \phi K^{\ast0}$ decay is given
as
\begin{equation}\label{fL-data}
 f_{L}= \left\{\begin{array}{ll}
 0.52\pm 0.05\pm 0.02 & \qquad {\rm BABAR}~\cite{babar1}, \\
 0.45\pm 0.05\pm 0.02 & \qquad {\rm Belle}~\cite{belle1}, \\
 0.57\pm 0.10\pm 0.05 & \qquad {\rm CDF}~\cite{cdf1}.
 \end{array}\right.
\end{equation}
On the other hand, since the two final-state light vector mesons
$\phi$ and ${K}^{*0}$ in this decay mode are flying out fleetly in
the rest frame of $B$ meson, and the structure of the charged weak
interaction current of the SM is left-handed, as well as the fact
that high-energy QCD interactions conserve helicity, any spin flip
of a fast flying quark will be suppressed by one power of $1/m_b$,
with $m_b$ the $b$ quark mass. It is therefore expected that, within
the SM, both of the final-state hadrons in this decay mode are
mainly longitudinally polarized, with
\begin{equation}\label{predifl}
f_{L}\sim 1-{\cal O}(1/m_b^2),
\end{equation}
while the transverse parts are suppressed by powers of $m_{\phi,
K^{*}}/m_{B}$. Obviously, the experimental data Eq.~(\ref{fL-data})
deviates significantly from the SM prediction Eq.~(\ref{predifl}),
and this polarization anomaly has attracted much interest in
searching for possible theoretical explanations both within the SM
and in various NP models~\cite{th_phik,kagan,KC,YY,kundu2}. For
example, the authors in Refs.~\cite{kagan,KC,YY,kundu2} have studied
this anomaly and found that the four-quark tensor operators of the
form $\bar{s}\sigma_{\mu\nu}(1+\gamma_{5})b\otimes
\bar{s}\sigma^{\mu\nu}(1+\gamma_{5})s$ could offer a simple
resolution to the observed polarization anomaly within some possible
parameter spaces.

Since the $B\rightarrow \eta K^{(\ast)}$ decays, in analogy with the
$B\rightarrow \phi K^{\ast}$ decays, also involve the $b\to
s\bar{s}s$ transition, it is necessary to investigate the effects of
these new types of four-quark tensor operators on the latter. In
particular, it is very interesting to see whether these new
four-quark tensor operators with the same parameter spaces could
also simultaneously account for the measured ${\cal B}(B\to\eta
K^{\ast})$~\cite{HFAG,Richichi:1999kj,babar2,belle2}, which are much
larger than the theoretical predictions within the
SM~\cite{Yang:2000xn,beneke1}, and survive bounds from $B\to\eta K$
decays. Motivated by these speculations, in this paper, we shall
investigate the effects of the following two types of tensor
operators on these decay modes~(with $i, j$ the color indices)
\begin{equation}\label{eq:tensor}
O_{T1}=\bar{s}\sigma_{\mu\nu}(1+\gamma_{5})b\otimes
\bar{s}\sigma^{\mu\nu}(1+\gamma_{5})s, \qquad
O_{T8}=\bar{s}_{i}\sigma_{\mu\nu}(1+\gamma_{5})b_{j}\otimes
\bar{s}_{j}\sigma^{\mu\nu}(1+\gamma_{5})s_{i},
\end{equation}
and try to find out the allowed parameter spaces characterized by
the strengths and phases of these new tensor operators that satisfy
all the experimental constraints from these decays. Moreover, since
the (pseudo-)scalar operators
\begin{equation}\label{eq:sp}
O_{S+P}=\bar{s}(1+\gamma_{5})b\otimes\bar{s}(1+\gamma_{5})s, \qquad
O^{\prime}_{S+P}=\bar{s}_{i}(1+\gamma_{5})b_{j}\otimes\bar{s}_{j}(1+\gamma_{5})s_{i}\,,
\end{equation}
can be expressed, through the Fierz transformations, as linear
combinations of the new tensor operators Eq.~(\ref{eq:tensor}),
constraints on these two operators can be derived easily from those
on the latter.

To further test such a particular NP scenario with anomalous tensor
operators, we also give predictions for the branching ratio and the
longitudinal polarization fraction of $B_{s}\to \phi \phi$ decay,
which is also involved the same quark level $b\to s\bar{s}s$
transition. All these results could be tested at the Fermilab
Tevatron and the  LHC-b experiments.

The paper is organized as follows. Sec.~2 is devoted to the
theoretical framework.  After a brief entertainment of the QCD
factorization formalism~(QCDF)~\cite{beneke2}, we discuss the
anomalous tensor operator contributions to the $B\to\eta K^{(\ast)}$
and $B\to\phi K^{\ast}$ decays. In Sec.~3, our numerical analysis
and discussions are presented. Sec.~4 contains our conclusions.
Appendix A recapitulates the amplitudes for the five decay modes
within the SM~\cite{Beneke:2006hg,beneke3}. All the theoretical
input parameters relevant to our analysis are summarized in Appendix
B.

\section{Theoretical Framework}

\subsection{The SM results within the QCDF framework}
In the SM, the effective Hamiltonian responsible for $b\to s$
transitions is given as~\cite{Buchalla:1996vs}
\begin{eqnarray}\label{eq:eff}
 {\cal H}_{\rm eff} &=& \frac{G_F}{\sqrt{2}} \biggl[V_{ub}
 V_{us}^* \left(C_1 O_1^u + C_2 O_2^u \right) + V_{cb} V_{cs}^* \left(C_1
 O_1^c + C_2 O_2^c \right) - V_{tb} V_{ts}^*\, \big(\sum_{i = 3}^{10}
 C_i O_i \big. \biggl. \nonumber\\
 && \biggl. \big. + C_{7\gamma} O_{7\gamma} + C_{8g} O_{8g}\big)\biggl] +
 {\rm h.c.},
\end{eqnarray}
where $V_{qb} V_{qs}^*$~($q=u, c$ and $t$) are products of the
Cabibbo-Kobayashi-Maskawa~(CKM) matrix elements~\cite{ckm}, $C_{i}$
the Wilson coefficients, and $O_i$ the relevant four-quark operators
whose explicit forms could be found, for example, in
Ref.~\cite{beneke2}.

In addition to ${\cal H}_{\rm eff}$, we must employ a factorisation
formalism of hadronic dynamics to study the $B^{-}\rightarrow \eta
K^{-}$, $\overline{B}^{0}\rightarrow \eta \overline{K}^{0}$,
$B^{-}\rightarrow \eta K^{\ast-}$, $\overline{B}^{0}\rightarrow \eta
\overline{K}^{\ast0}$, and $\overline{B}^{0}\rightarrow \phi
\overline{K}^{\ast0}$ decays. To this end, we take the framework of
QCDF~\cite{beneke2}. The factorization formula allows us to
calculate systemically the hadronic matrix element of operators in
the effective Hamiltonian Eq.~(\ref{eq:eff})
\begin{eqnarray}\label{eq:QCDF}
\langle M_{1}M_{2}|O_i|B\rangle&=&\sum_{j}F_{j}^{B\rightarrow
M_{1}}\int_{0}^{1}dx
T_{ij}^{I}(x)\Phi_{M_2}(x)+(M_1\leftrightarrow M_2) \nonumber \\
&   &+\int_{0}^{1}d\xi\int_{0}^{1}dx\int_{0}^{1}dy
T_{i}^{II}(\xi,x,y)\Phi_{B}(\xi)\Phi_{M_1}(x)\Phi_{M_2}(y),
\end{eqnarray}
where $F_{j}^{B\rightarrow M}$ is the $B\rightarrow M$ transition
form factor, $T_{ij}^{I}$ and $T_{i}^{II}$ are the perturbatively
calculable hard kernels, and $\Phi_{X}(x)~(X=B,M_{1,2})$ are the
universal nonperturbative light-cone distribution amplitudes~(LCDAs)
of the corresponding mesons.

In the recent years, the QCDF formalism has been employed
extensively to study non-leptonic $B$ decays. For example, all the
decay modes considered here have been studied comprehensively within
the SM in Refs.~\cite{Beneke:2006hg,beneke3}. We recapitulate the
amplitudes for $B\to\eta K^{(\ast)}$ and $B\to\phi K^{\ast}$ decays
in Appendix A.

It is noted that, along with its many novel progresses in
non-leptonic $B$ decays, the framework contains estimates of some
power corrections which can not be computed rigorously. These
contributions may be numerically important for realistic $B$-meson
decays, especially for some penguin-dominated decay
modes~\cite{Beneke:2006hg,kagan,beneke3,Li:2006jb}. In fact, there
are no reliable methods available at present to \emph{calculate}
such contributions. To give \emph{conservative} theoretical
predictions, at least, one should leave the associated parameters
varying in reasonable regions to show their possible effects. In this work,
following closely the treatment made in
Refs.~\cite{beneke3,Feldmann:2004mg}, we will parameterize the
end-point divergences associated with these power corrections as
 \begin{equation}\label{treat-for-anni}
\int_0^1 \frac{\!dx}{x}\, \to X_A =e^{i\phi_A} \ln
\frac{m_B}{\Lambda_h}, \qquad \int_0^1 \frac{\!dx}{x^2}\,
\to X_L = \frac{m_B}{\Lambda_h} e^{i\phi_A}-1.
 \end{equation}
In the following numerical calculations, we take the parameter
$\Lambda_h$ and the phase $\phi_A$ varying in the range
$0.2\sim0.8~{\rm GeV}$ and $-45^{\circ}\sim45^{\circ}$,
respectively.

In our calculation, we have neglected possible intrinsic charm
content and anomalous gluon couplings related to the meson $\eta$,
both of which have been shown to have only marginal effects on the
four $B\to \eta K^{(\ast)}$ decays~\cite{Yang:2000xn,beneke1}. As
for the $\eta$--$\eta'$ mixing effects, we shall adopt the
Feldmann-Kroll-Stech~(FKS) scheme~\cite{FKS} as implemented in
Ref.~\cite{beneke1}. A recent study and comparison of different
$\eta$--$\eta'$ mixing schemes has been given in Ref.~\cite{Frere}.

In the amplitude for $\bar{B}^{0}\to \phi \bar{K}^{\ast}$ decay in
Eq.~(\ref{eq:phiksm}), a new power-enhanced electromagnetic penguin
contribution to the negative-helicity electroweak penguin
coefficient $\alpha_{3,{\rm EW}}^{p,-}$, as first noted by Beneke,
Rohrer and Yang~\cite{Beneke:2006hg}, has also been taken into
account in our calculation. In a recent comprehensive study of $B\to
VV$ decays~\cite{BenVV}, it is found that the small $f_{L}$ could be
accommodated within the SM, however, with very large theoretical
uncertainties.

\subsection{Anomalous tensor operators and their contributions to the decay amplitudes}

Since the SM may have  difficulties in explaining the large ${\cal
B}(B\rightarrow \eta K^{\ast})$ and the measured polarization
observables in $B\rightarrow \phi K^{\ast}$ decays, we shall discuss
possible NP resolutions to these observed discrepancies.
Specifically, we shall investigate whether these discrepancies could
be resolved by introducing two anomalous four-quark tensor operators
defined by Eq.~(\ref{eq:tensor}).

We write the NP effective Hamiltonian as
\begin{equation}\label{eq:np}
{\cal H}_{eff}^{\rm NP}=\frac{G_{F}}{\sqrt{2}}\, |V_{ts}|\,
e^{i\delta_{T}}\,\Big[C_{T1}O_{T1} +C_{T8}O_{T8}\Big]+ {\rm h.c.},
\end{equation}
with the tensor operators $O_{T1}$ and $O_{T8}$ defined by
Eq.~(\ref{eq:tensor}). The coefficient $C_{T1(T8)}$ describes the
relative interaction strength of the tensor operator $O_{T1(T8)}$,
and $\delta_{T}$ is the NP weak phase. In principle, such four-quark
tensor operators could be produced in various NP scenarios, e.g., in
the Minimal Supersymmetric Standard Model~(MSSM)~\cite{csh,hiller}.
Interestingly, the recent study of radiative pion decay $\pi^+ \to
e^+ \nu \gamma$ at PIBETA detector~\cite{pibeta} has found
deviations from the SM predictions in the
high-$E_{\gamma}$--low-$E_{e^+}$ kinematic region, which may
indicate the existence of anomalous tensor quark-lepton
interactions~\cite{pob,chizhov,Voloshin:1992sn}.

At first, we present the NP contributions to the decay amplitudes of
$B\to \eta K^{(\ast)}$ and $B\to \phi K^{\ast}$ decays due to these
new tensor operators. Since their coefficients are unknown parameters,
 for simplicity, we shall only consider the
leading contributions of these tensor operators.

%%%%%%%%%%%%%%%%%%%%%%%%%%%%%%%%%%%%%%%%%%%%%%%%%%
\begin{figure}[ht]
\epsfxsize=9cm \centerline{\epsffile{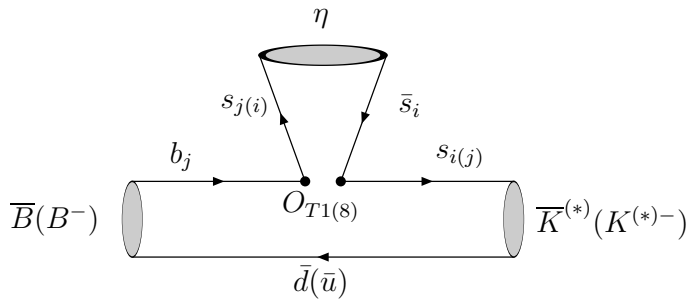}}
\centerline{\parbox{16cm}{\caption{\label{fig1} \small Feynman
diagram contributing to the decay amplitudes of $B\rightarrow\eta
K^{(\ast)}$ decays due to the anomalous tensor operators. Another
type of insertion has no contribution. }}}
\end{figure}
%%%%%%%%%%%%%%%%%%%%%%%%%%%%%%%%%%%%%%%%%%%%%%%%%%%%

As for the four $B\rightarrow \eta K^{(\ast)}$ decays, the relevant
Feynman diagram due to the tensor operators $O_{T1,T8}$ is shown in
Fig.~\ref{fig1}. It is easy to realize that the amplitude
corresponding to another type of insertion would vanish in the
leading order approximation. Instead of using the Fierz
transformation, an easy way to calculate the amplitude in
Fig.~\ref{fig1} is to use the light-cone projection operator of the
meson $\eta$ in momentum space~\cite{beneke3}
\begin{equation}\label{projector}
M^{\eta_s}_{\alpha\beta}= \frac{i f_{\eta}^{s}}{4} \left[\spur{q}
\gamma_{5}\, \Phi_{\eta}(x) - \mu_{\eta_s} \gamma_{5}\,
\frac{\spur{k_2}\spur{k_1}}{k_{2}\cdot k_{1}}\, \Phi_{p}(x)
\right]_{\alpha\beta},
\end{equation}
where $q$, $\Phi_{\eta}$, and $\Phi_{p}$ are the momentum,
leading-twist, and twist-3 LCDAs of the meson $\eta$, respectively.
$k_{1}^{\mu}$ and $k_{2}^{\mu}$ denote the momenta of the quark and
anti-quark in the meson $\eta$, and are given by
\begin{equation}
k_{1}^{\mu}=x q^{\mu} + k_{\perp}^{\mu} +
\frac{\vec{k}_{\perp}^{2}}{2x q\cdot \bar{q}}\, \bar{q}^{\mu},
\qquad k_{2}^{\mu}=\bar{x} q^{\mu} - k_{\perp}^{\mu} +
\frac{\vec{k}_{\perp}^{2}}{2\bar{x} q\cdot \bar{q}}\, \bar{q}^{\mu},
\end{equation}
with $\bar{x}=1-x$, and $x$ the momentum fraction carried by the
constituent quark. The decay constant $f_{\eta}^s$ and the factor
$\mu_{\eta_s}$ in Eq.~(\ref{projector}) are defined, respectively,
by~\cite{beneke1}
\begin{equation}
\langle\eta(q)|\bar{s}\gamma_{\mu} \gamma_{5} s|0\rangle = - i
f_{\eta}^{s} q_{\mu}, \qquad \mu_{\eta_s} =
\frac{\overline{m}_{b}}{2}\, r_{\chi}^{\eta_s}=\frac{h_{\eta}^{s}}{2
f_{\eta}^{s}\, \overline{m}_{s}},
\end{equation}
where we have used $|\eta\rangle=\cos \phi|\eta_{q}\rangle
-\sin\phi|\eta_{s}\rangle$, with $|\eta_q \rangle
=(|\bar{u}u\rangle+|\bar{d}d\rangle)/\sqrt{2}$ and $ |\eta_s \rangle
=|\bar{s}s\rangle$~\cite{FKS}.

After some simple calculations, the NP contributions to the decay
amplitudes of the four $B\rightarrow \eta K^{(\ast)}$ decays due to
${\mathcal H}^{NP}_{eff}$  in Eq.~(\ref{eq:np}) can be written as
\begin{eqnarray}
{\cal A}^{\rm NP}_{B^{-}\rightarrow\eta
K^{-}}&=&i\,\frac{G_{F}}{\sqrt{2}}\, |V_{ts}|\, e^{i\delta_{T}} 3\,
g_{T}
r_{\chi}^{\eta_s}\,\left(m_{B_u}^2-m_{K^-}^2\right)\,F_{0}^{B\rightarrow
K}(m_{\eta}^2)\,f_{\eta}^s\,,\label{amp1_NP}\\
{\cal A}^{\rm NP}_{\overline{B}^{0}\rightarrow\eta
\overline{K}^{0}}&=&i\,\frac{G_{F}}{\sqrt{2}}\, |V_{ts}|\,
e^{i\delta_{T}} 3\, g_{T}\,
r_{\chi}^{\eta_s}\,\left(m_{B_d}^2-m_{K^0}^2\right)\,F_{0}^{B\rightarrow
K}(m_{\eta}^2)\,f_{\eta}^s\,,\label{amp2_NP}\\
{\cal A}^{\rm NP}_{B^{-}\rightarrow\eta K^{\ast-}}&=&-i\,\sqrt{2}\,
G_{F}\, |V_{ts}|\, e^{i\delta_{T}} 3\, g_{T}\,
 r_{\chi}^{\eta_s}\,
m_{K^{\ast-}}\,(\varepsilon_{2}^{\ast}\cdot p)\,A_{0}^{B\rightarrow
K^{\ast}}(m_{\eta}^2)\,f_{\eta}^s\,,\label{amp3_NP}\\
{\cal A}^{\rm NP}_{\overline{B}^{0}\rightarrow\eta
\overline{K}^{\ast 0}}&=&-i\,\sqrt{2}\, G_{F}\, |V_{ts}|\,
e^{i\delta_{T}} 3\, g_{T}\, r_{\chi}^{\eta_s}\, m_{K^{\ast
0}}\,(\varepsilon_{2}^{\ast}\cdot p)\,A_{0}^{B\rightarrow
K^{\ast}}(m_{\eta}^2)\,f_{\eta}^s\,\label{amp4_NP},
\end{eqnarray}
where $g_{T}=C_{T8}+C_{T1}/N_{c}$, and the factor 3 is due to
contractions of the involved $\gamma$ matrices. It is interesting to
note that the above four decay amplitudes are all proportional to
the ``chirally-enhanced" factor
$r_{\chi}^{\eta_s}=\frac{h_{\eta}^{s}}{
f_{\eta}^{s}\,\overline{m}_{b}\, \overline{m}_{s}}$, which has been
found to be very important for charmless hadronic $B$
decays~\cite{beneke3}.

%%%%%%%%%%%%%%%%%%%%%%%%%%%%%%%%%%%%%%%%%%%%%%%%%%
\begin{figure}[ht]
\epsfxsize=13cm \centerline{\epsffile{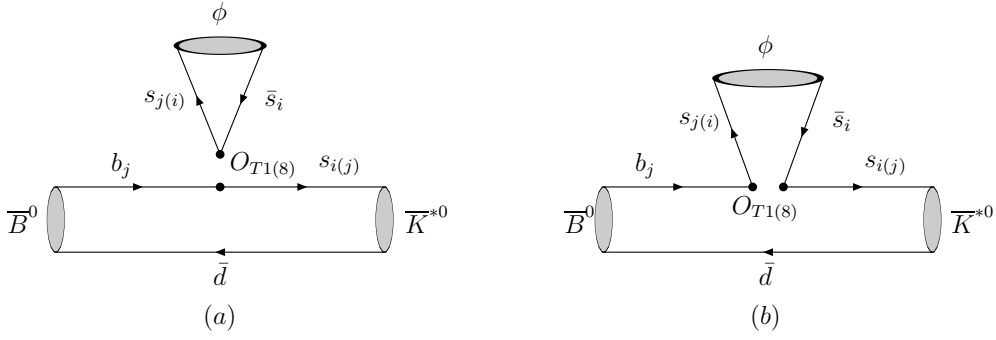}}
\centerline{\parbox{16cm}{\caption{\label{fig2} \small Feynman
diagrams contributing to the decay amplitude of
$\overline{B}^{0}\rightarrow\phi{\overline{K}}^{\ast0}$ decay due to
the anomalous tensor operators.}}}
\end{figure}
%%%%%%%%%%%%%%%%%%%%%%%%%%%%%%%%%%%%%%%%%%%%%%%%%

We now present the NP contribution to the decay amplitude of
$\overline{B}^{0}\rightarrow\phi{\overline{K}}^{\ast0}$ decay. Based
on the observation that they contribute only to the transverse
polarization amplitudes but not to the longitudinal
one~\cite{kagan,KC,YY}, these anomalous four-quark tensor operators
have been proposed to resolve the polarization anomaly observed in
$B\rightarrow\phi K^{\ast}$ decays. The relevant Feynman diagrams
are shown in Fig.~\ref {fig2}. For Fig.~\ref{fig2}~(a), we shall use
the following matrix elements~\cite{projector,BSW,formfactor}
\begin{eqnarray}
\langle\phi(q,\varepsilon_{1})|\bar{s}\sigma^{\mu\nu}s|0\rangle &=&
-f^{T}_{\phi} (\varepsilon_{1\perp}^{\mu\ast} q_{\nu} -
\varepsilon_{1\perp}^{\nu\ast} q_{\mu})\,,\\
\langle \overline K^*(p^{\prime},\varepsilon_{2})|\bar{s}
\sigma_{\mu\nu} q^\nu b |\overline B(p)\rangle &=&
\epsilon_{\mu\nu\rho\sigma} \varepsilon_{2}^{*\nu} p^\rho
p^{\prime\sigma} \, 2 T_1^{B\rightarrow K^{\ast}}(q^2)\,, \nonumber\\
\langle \overline K^*(p^{\prime},\varepsilon_{2})|\bar{s}
\sigma_{\mu\nu} q^\nu \gamma_5 b |\overline B(p)\rangle &=& (-i)
T_2^{B\rightarrow K^{\ast}}(q^2) \Big\{ \varepsilon^*_{2,\mu}
(m_B^2-m_{K^*}^2)
- (\varepsilon^{*}_{2}\cdot p) \,(p + p^{\prime})_\mu \Big\} \nonumber\\
&+& (-i) T_3^{B\rightarrow K^{\ast}}(q^2)(\varepsilon_2^* \cdot p)
\Big\{ q_\mu - \frac{q^2}{m_B^2-m_{K^*}^2}\, (p + p^{\prime})_\mu
\Big\}.
\end{eqnarray}
For Fig.~\ref{fig2}~(b), we shall use the light-cone projector
operator of the transversely polarized vector meson
$\phi$~\cite{Beneke:2006hg,projector}
\begin{equation}
M_{\perp}^{\phi} = -\frac{i f^{T}_{\phi}}{4}
\spur{\varepsilon}^{\ast}_{1\perp} \spur{q} \Phi_{\perp}(x) +
\cdots,
\end{equation}
where $\Phi_{\perp}(x)$ is the leading twist LCDA of the meson
$\phi$, and the ellipsis denotes additional parts that have no
contributions in our case. It is easy to find that neither
Fig.~\ref{fig2}(a) nor (b) contributes to the longitudinal
polarization amplitude, and the final decay amplitude
 can be written as
\begin{eqnarray}\label{phik_NP}
{\cal A}_{\overline{B}^{0}\rightarrow \phi
\overline{K}^{\ast0}}^{\rm NP} &=& \frac{G_{F}}{\sqrt{2}}\,
|V_{ts}|\, e^{i\delta_{T}} g^{\prime}_{T}
 (-4if_{\phi}^{T})\, \biggl\{i
\epsilon_{\mu\nu\rho\sigma}\varepsilon_{1\perp}^{\ast\mu}
\varepsilon_{2}^{\ast \nu} p^{\rho} p^{\prime\sigma}
2 T_{1}^{B\rightarrow K^{\ast}}(m_{\phi}^{2}) \, \biggl. \nonumber\\
&& \biggl. + T_{2}^{B\rightarrow K^{\ast}}(m_{\phi}^2)
\Big[(\varepsilon_{1\perp}^{\ast} \cdot
\varepsilon_{2}^{\ast})(m_{B}^{2} - m_{K^{\ast}}^{2}) -
2(\varepsilon_{1\perp}^{\ast} \cdot p)(\varepsilon_{2}^{\ast} \cdot
p)
\Big]\, \biggl. \nonumber\\
&& \biggl. -2 T_{3}^{B\rightarrow
K^{\ast}}(m_{\phi}^2)\,\frac{m_{\phi}^2}{m_{B}^{2} -
m_{K^{\ast}}^{2}}\, (\varepsilon_{1\perp}^{\ast} \cdot p) (
\varepsilon_{2}^{\ast}\cdot p) \biggl\}\,,
\end{eqnarray}
with $g^{\prime}_{T}=(1+\frac{1}{2 N_{c}})\,C_{T1}+(
\frac{1}{2}+\frac{1}{N_{c}})\,C_{T8}$. In the helicity basis, the
new decay amplitude Eq.~(\ref{phik_NP}) can be further decomposed
into
\begin{eqnarray}
H^{\rm NP}_{00} &=& 0, \label{hphi1_np}\\
H^{\rm NP}_{\pm\pm} &=& \frac{G_{F}}{\sqrt{2}}\, |V_{ts}|\,
e^{i\delta_{T}} g^{\prime}_{T} (4 i
f_{\phi}^{T})\bigg[(m_{B_{d}}^{2} -
m_{K^{\ast0}}^{2})\,T_{2}^{B\rightarrow K^{\ast}}(m_{\phi}^2) \mp 2
m_{B_{d}} p_{c}\, T_{1}^{B\rightarrow
K^{\ast}}(m_{\phi}^2)\bigg]\,,\label{hphi2_np}
\end{eqnarray}
where $p_{c}$ is the center-of-mass momentum of final mesons in
$\bar{B}^0$ rest frame. Compared with the SM predictions
Eqs.~(\ref{hphi1_SM}) and (\ref{hphi2_SM}), the new transverse
polarization amplitudes $H^{\rm NP}_{\pm\pm}$ are enhanced by a
factor of $m_{B_d}/m_{\phi}$, while the longitudinal part remains
unchanged. It is therefore expected that these new tensor operators
might provide a possible resolution to the polarization anomaly
observed in $B\to \phi K^{\ast}$ decays.

\subsection{The branching ratios and polarization fractions}

From the above discussions, the total decay amplitudes are then
given as
\begin{equation}\label{totalamp}
\mathcal{A}=\mathcal{A}^{\rm SM} + \mathcal{A}^{\rm NP},
\end{equation}
where $\mathcal{A}^{\rm SM}$ denotes the SM results obtained using
the QCDF, and $\mathcal{A}^{\rm NP}$ the contributions of the
particular NP scenario with anomalous tensor operators in
Eq.~(\ref{eq:np}). The corresponding branching ratios are
\begin{eqnarray}
{\mathcal B}(B^{0,+}\rightarrow\eta K^{(\ast) 0,+}) &=&
\frac{\tau_{B} \,p_{c}}{8 \pi m_{B}^{2}}\,
|\mathcal{A}(B^{0,+}\rightarrow\eta K^{(\ast)0,+})|^{2}, \\
{\mathcal B}(\overline{B}^{0}\rightarrow\phi\overline{K}^{\ast0})
&=&
\frac{\tau_{B}\,p_{c}}{8 \pi m_{B}^{2}}\, (|H_{00}|^{2}+
|H_{++}|^{2}+ |H_{--}|^{2})\,,
\end{eqnarray}
where $\tau_{B}$ is the life time of $B$ meson.

In the transversity basis~\cite{transversebasis}, the decay
amplitude for any $\overline{B}\rightarrow V V$ decay can also be
decomposed into another three quantities $A_0$, $A_{\parallel}$, and
$A_{\perp}$, which are related to the helicity amplitudes $H_{00}$,
$H_{++}$, and $H_{--}$ through
\begin{equation}
A_{0}=H_{00}, \qquad A_{\parallel}=\frac{H_{++}+H_{--}}{\sqrt2},
\qquad A_{\perp}=-\frac{H_{++}-H_{--}}{\sqrt2}\,.
\end{equation}
In terms of these quantities, we can express the longitudinal
polarization fraction as
\begin{eqnarray}
f_{L}&=&\frac{|A_{0}|^{2}}{|A_{0}|^{2} + |A_{\parallel}|^{2} +
|A_{\perp}|^{2}}\, \nonumber \\
&=& \frac{|H_{00}|^{2}}{|H_{00}|^{2}+
|H_{++}|^{2}+ |H_{--}|^{2}}\,.
\end{eqnarray}
In addition, the relative phases between these helicity amplitudes
$\phi_{\parallel,\perp}={\rm
Arg}(A_{\parallel,\perp}/A_{0})+\pi$~(the definition of these
observables is compatible with that used by the BABAR and Belle
collaborations~\cite{babar1,belle1}), are potentially very useful
for constraining the parameter spaces of NP scenario, however,
depend on whether the strong phases of the helicity amplitudes could
be calculated reliably.

\section{Numerical analysis and discussions}
With the theoretical formulas and the input parameters summarized in
Appendix B for the decay modes of our concerns, we now go to our
numerical analysis and discussions.

As shown by the NP decay amplitudes
Eqs.~(\ref{amp1_NP})--(\ref{amp4_NP}) and
(\ref{hphi1_np})--(\ref{hphi2_np}), the allowed regions for the
parameters $g_{T}$ and $\delta_{T}$ can be obtained from the four
measured ${\mathcal B}(B^{0,+}\rightarrow\eta K^{(\ast)0,+})$, and
the ones for $g^{\prime}_{T}$ from ${\mathcal
B}(B^{0}\rightarrow\phi K^{\ast0}$) and $f_{L}$, respectively.
Generally, we have five branching ratios and one polarization
fraction, but only three free parameters: one NP weak phase
$\delta_{T}$ and two effective coefficients $g_{T}$ and
$g^{\prime}_{T}$~(or equivalently $C_{T1}$ and $C_{T8}$). So, it is
easy to guess that these decays could severely constrain or rule out
the NP scenario with two anomalous tensor operators
Eq.~(\ref{eq:np}).

To make the guess clear, we shall first find out the allowed regions
for the parameters $g_{T}$ and $\delta_{T}$ from the four
$B^{0,+}\rightarrow\eta K^{(\ast)0,+}$ decays. Then, using the
allowed regions for the weak phase $\delta_{T}$, we try to put
constraint on the parameter $g^{\prime}_{T}$  from  ${\mathcal
B}(B^{0}\rightarrow\phi K^{\ast0})$ and $f_{L}$. Finally, we can
obtain the allowed parameter spaces, if there are, for $C_{T1}$,
$C_{T8}$, and $\delta_{T}$ that satisfy all the experimental data on
the decay modes of our concerns. To further test the particular NP
scenario Eq.~(\ref{eq:np}), we also present our theoretical
predictions for $B_{s}\rightarrow\phi\phi$ decay. Theoretical
estimations of the annihilation contributions at present suffer from
very large uncertainties, which of course will dilute the
requirement of NP very much. To show the dilution, we will give our
numerical results for two cases, i.e., with and without
annihilations for comparison.

\subsection{Constraints on the NP parameters from $B^{0,+}\rightarrow\eta K^{(\ast)0,+}$ decays}

\begin{figure}[t]
\begin{center}
\epsfxsize=15cm {\epsffile{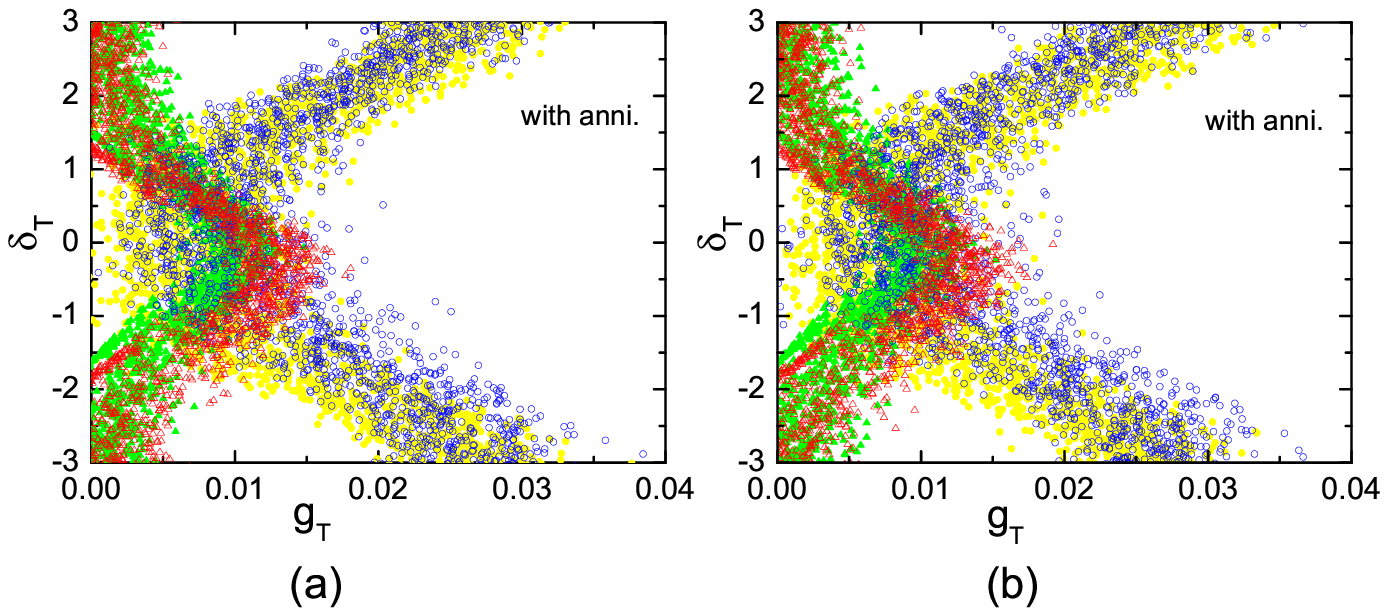}}\\
\epsfxsize=15cm {\epsffile{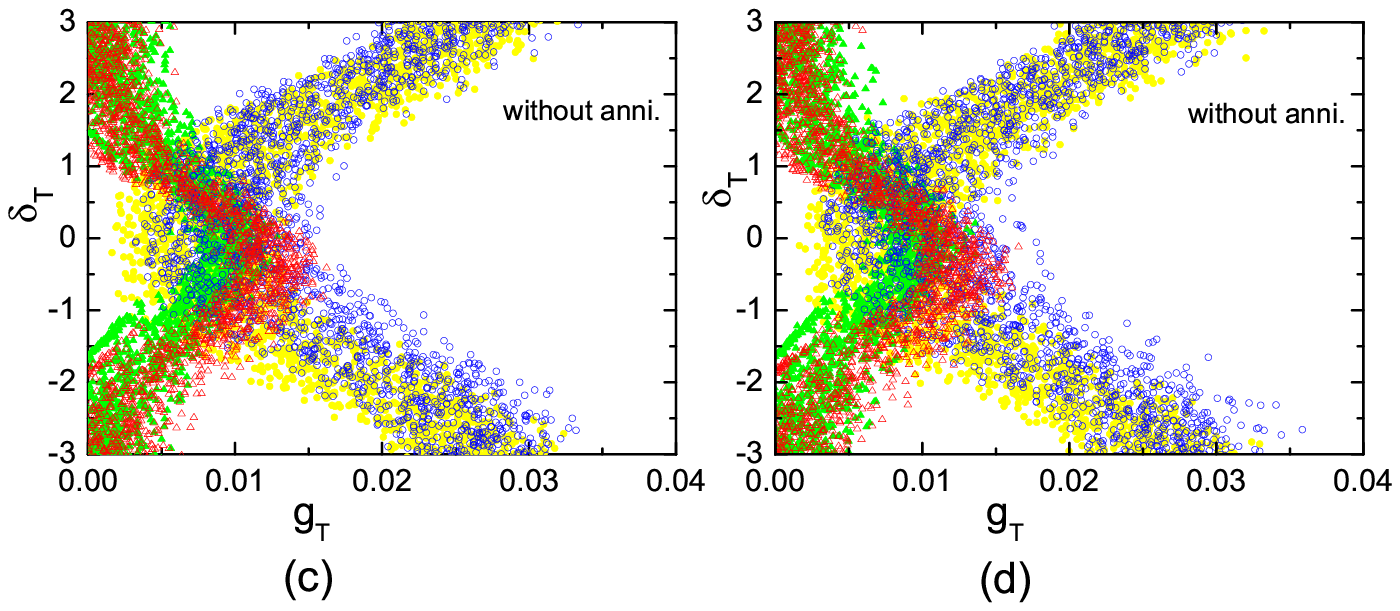}}
\centerline{\parbox{16cm}{\caption{\label{gtdeltat1} \small The
contour plots for the parameters $g_{T}$ and $\delta_{T}$ with the
experimental data varying within $1\sigma$~((a) and (c)) and
$2\sigma$~((b) and (d)) error bars, respectively. The red triangle,
green triangle, blue circle, and yellow circle bands come from the
decays $B^{-}\rightarrow\eta K^{-}$, $B^{0}\rightarrow\eta K^{0}$,
$B^{-}\rightarrow\eta K^{\ast,-}$, and $B^{0}\rightarrow\eta
K^{\ast0}$, respectively.  Plots labels `with(out) anni.' denote the
results with(out) the annihilation contributions. }}}
\end{center}
\end{figure}

\begin{figure}[t]
\begin{center}
\epsfxsize=15cm {\epsffile{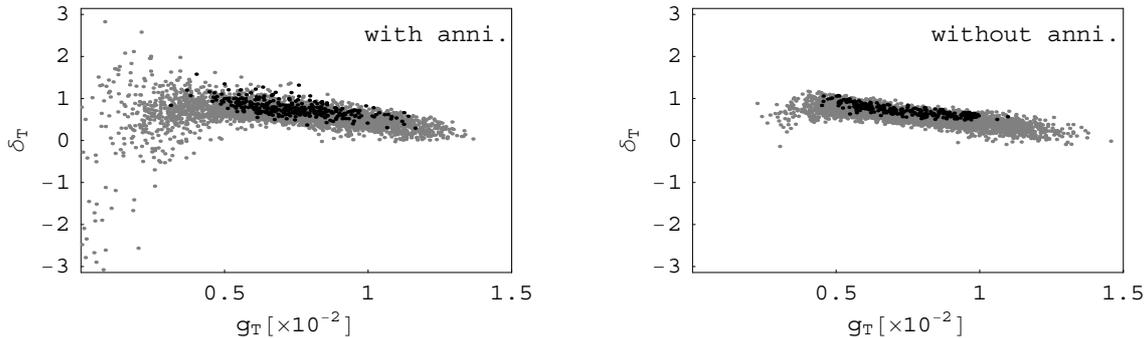}}
\centerline{\parbox{16cm}{\caption{\label{gtdeltat2} \small The
allowed regions for the parameters $g_{T}$ and $\delta_{T}$
constrained by the four $B^{0,+}\rightarrow\eta K^{(\ast)0,+}$
decays. The dark and the gray regions correspond to the results
obtained with the measured branching ratios varying within $1\sigma$
and $2\sigma$ error bars, respectively. The left~(right) plot
denotes the results with(out) the annihilation contributions. }}}
\end{center}
\end{figure}

Since the tensor operator contributions to the decay amplitudes of
the four $B^{0,+}\rightarrow\eta K^{(\ast)0,+}$ decays are all
characterized by the parameters $g_{T}$ and $\delta_{T}$, possible
regions for these two NP parameters can be obtained from the
measured branching ratios of $B^{0,+}\rightarrow\eta K^{(\ast)0,+}$
decays. Our main results are shown in
Tables~\ref{br1-4comparison}--\ref{gtdeltatresult} and
Figs.~\ref{gtdeltat1}--\ref{gtdeltat2}. The experimental data listed
in Table~\ref{br1-4comparison} are taken from the Heavy Flavor
Averaging Group~(HFAG)~\cite{HFAG}. The SM predictions for the
branching ratios of these four decays are presented in the third
column of Table~\ref{br1-4comparison}, where the theoretical
uncertainties are obtained by varying the input parameters within
the regions specified in Eq.~(\ref{treat-for-anni}) and Appendix B.
For each decay mode, the first and the second row are evaluated with
and without the annihilation contributions, respectively.

%%%%%%%%%%%%%%%%%%%%%%%%%%%%%%%%%%%%%%%%
\begin{table}[ht]
\centerline{\parbox{16cm} {\caption{\label{br1-4comparison} \small
Experimental data~\cite{HFAG} and theoretical predictions for the
branching ratios~(in units of $10^{-6}$). The numbers in columns
Case I and Case II are our fitting results with $g_{T}$ and
$\delta_{T}$ constrained by varying the experimental data within
$1\sigma$ and $2\sigma$ error bars, respectively. For each decay
mode, the first~(second) row is evaluated with(out) the annihilation
contributions. }}}
\begin{center}
\begin{tabular}{lcccc}\hline\hline
Decay channel & Experiment & SM & Case I & Case II
\\\hline
$B^{-}\rightarrow\eta K^{-}$
  &$2.2\pm0.3$&$2.21^{+1.35}_{-0.85}$&$2.29^{+0.12}_{-0.18}$&$2.19^{+0.35}_{-0.33}$\\
$$&$$&$1.80\pm0.82$&$2.19\pm0.27$&$2.17\pm0.34$\\
${B}^{0}\rightarrow\eta{K}^{0}$
  &$<1.9$&$1.34^{+1.09}_{-0.65}$&$1.30^{+0.14}_{-0.14}$&$1.25^{+0.31}_{-0.28}$\\
$$&$$&$1.02\pm0.65$&$1.42\pm0.22$&$1.34\pm0.30$\\
$B^{-}\rightarrow\eta K^{\ast-}$
  &$19.5^{+1.6}_{-1.5}$&$6.27^{+3.4}_{-2.3}$&$18.15^{+0.21}_{-0.09}$&$17.44^{+0.68}_{-0.55}$\\
$$&$$&$5.03\pm1.82$&$17.88\pm0.49$&$17.45\pm0.61$\\
${B}^{0}\rightarrow\eta{K}^{\ast0}$
  &$16.1\pm1.0$&$6.87^{+3.5}_{-2.5}$&$16.96^{+0.09}_{-0.16}$&$17.13^{+0.56}_{-0.68}$\\
$$&$$&$5.60\pm1.95$&$17.10\pm0.29$&$17.03\pm0.65$
\\\hline \hline
\end{tabular}
\end{center}
\end{table}
%%%%%%%%%%%%%%%%%%%%%%%%%%%%%%%%%%%%%%%%%%%%

From Table~\ref{br1-4comparison}, we can see that the theoretical
predictions within the SM for both ${\mathcal B}(B^{-}\to \eta K^{-}
)$ and ${\mathcal B}(B^{0}\to \eta K^{0} )$ agree with the
experimental data within errors. However, both ${\mathcal B}(
B^{-}\to \eta K^{\ast-})$ and ${\mathcal B}( B^{0}\to \eta
K^{\ast0})$ are quite lower than the experimental data. We also note
that our results are a little different with those in
Refs.~\cite{Yang:2000xn,beneke1,beneke3}, due to different choices
for the input parameters, such as the moderate strength of $X_{A}$,
$\lambda_{B}$, and so on.

As shown in Fig.~\ref{gtdeltat1}, the four ${\mathcal
B}(B^{0,+}\rightarrow\eta K^{(\ast)0,+})$ are very sensitive to the
presence of the ${\mathcal H}^{NP}_{eff}$ of Eq.~(\ref{eq:np}). The
two bands for the parameter spaces constrained by $B^{-}\to\eta
K^{-}$ and $B^{0}\to \eta K^{0}$ decays are much overlapped, and the
same situation is also found for the two bands constrained by
$B^{-}\to \eta K^{\ast-}$ and $B^{0}\to \eta K^{\ast0}$ decays.
However, we note that ${\mathcal B}(B\to \eta K)$ and ${\mathcal
B}(B\to \eta K^{\ast})$ have quite different dependence on these NP
contributions. So, the allowed regions for the parameters $g_{T}$
and $\delta_{T}$  are severely narrowed down when constraints from
these four decay modes are combined.

%%%%%%%%%%%%%%%%%%%%%%%%%%%%%%%%%%%%%%%%%%%%%%%%%%
%%%%%%%%%%%%%%%%%%%%%%%%%%%%%%%%%%%%%%%%%%%%%%%%%%

The final allowed regions for the parameters $g_{T}$ and
$\delta_{T}$ extracted from the four $B^{0,+}\rightarrow\eta
K^{(\ast)0,+}$ decays are shown in Fig.~\ref{gtdeltat2}, where the
left~(right) plot is the results with(out) the annihilation
contributions. In addition, the dark and the gray regions in
Fig.~\ref{gtdeltat2} correspond to the results obtained with the
measured branching ratios varying within $1\sigma$ and $2\sigma$
error bars, respectively. From now on, we denote these two possible
regions by Case I and Case II. Comparing the two plots in
Fig.~\ref{gtdeltat2}, one can find that the uncertainties of
annihilation contributions would loosen constraints on the NP
parameters, especially on $\delta_{T}$. The numerical results for
the parameters $g_{T}$ and $\delta_{T}$ corresponding to the above
two allowed regions are presented in Table~\ref{gtdeltatresult}.

As shown in Table~\ref{br1-4comparison}, with the parameters $g_{T}$
and $\delta_{T}$ varying within these two allowed regions, the large
${\mathcal B}(B^{-}\to \eta K^{\ast-})$ and ${\mathcal B}(B^{0}\to
\eta K^{\ast0})$ can be accounted for by the anomalous four-quark
tensor operators without violating ${\mathcal B}(B^{-}\to \eta
K^{-})$. It is also interesting to note that, taking the $90\%$ CL
upper limit of ${\mathcal B} (B^{0}\rightarrow\eta{K}^{0})$ as an
input, our fitting result for ${\mathcal
B}(B^{0}\rightarrow\eta{K}^{0})$ is in good agreement with the very
recent measurements
\begin{eqnarray}
&&{\mathcal
B}(B^{0}\to\eta{K}^{0})=(1.8^{+0.7}_{-0.6}\pm0.1)\times10^{-6}
\qquad {\rm BABAR}~\cite{babar2}\,, \\
&&{\mathcal B}(B^{0}\to \eta{K}^{0})=(1.1\pm0.4\pm0.1)\times10^{-6}
\qquad {\rm Belle}~\cite{belle2}\,,
\end{eqnarray}
which give the average value ${\mathcal B}
(B^{0}\rightarrow\eta{K}^{0})=(1.3\pm0.3)\times10^{-6}$.

%%%%%%%%%%%%%%%%%%%%%%%%%%%%%%%%%%%%%%%%%%%%%%%%%%
%%%%%%%%%%%%%%%%%%%%%%%%%%%%%%%%%%%%%%%%%%%%%%%%%

%%%%%%%%%%%%%%%%%%%%%%%%%%%%%%%%%%%%%%%%%%%%%%%%%
\begin{table}[ht]
\centerline{\parbox{16cm} {\caption{\label{gtdeltatresult} \small
The numerical results for the parameters $g_{T}$ and $\delta_{T}$,
corresponding to the two allowed regions shown in
Fig.~\ref{gtdeltat2}. The allowed regions for the parameter
$g_{T}^{\prime}$ constrained by $B^{0}\to\phi K^{\ast0}$ decay are
also presented. For each case, the first~(second) row denotes the
results with(out) the annihilation contributions. }}}
\begin{center}
\begin{tabular}{lcccc}\hline\hline
& Allowed region & $g_{T}(\times10^{-3})$ & $g'_{T}(\times10^{-3})$
& $\delta_{T}({\rm rad})$ \\ \hline
Case I &  dark&$7.3^{+1.6}_{-1.5}$&$7.6^{+1.0}_{-0.9}$&$0.77^{+0.20}_{-0.16}$\\
&&$7.7\pm1.6$&$8.9\pm0.5$&$0.70\pm0.13$\\
Case II & gray &$7.2^{+2.5}_{-2.8}$&$8.3^{+1.3}_{-1.2}$&$0.60^{+0.29}_{-0.39}$\\
&&$8.0\pm2.3$&$9.1\pm0.6$&$0.55\pm0.20$
\\\hline \hline
\end{tabular}
\end{center}
\end{table}
%%%%%%%%%%%%%%%%%%%%%%%%%%%%%%%%%%%%%%%

\subsection{Constraints on the NP parameters from $B^{0}\rightarrow\phi K^{\ast0}$ decay}

\begin{figure}[ht]
\begin{center}
\epsfxsize=14cm {\epsffile{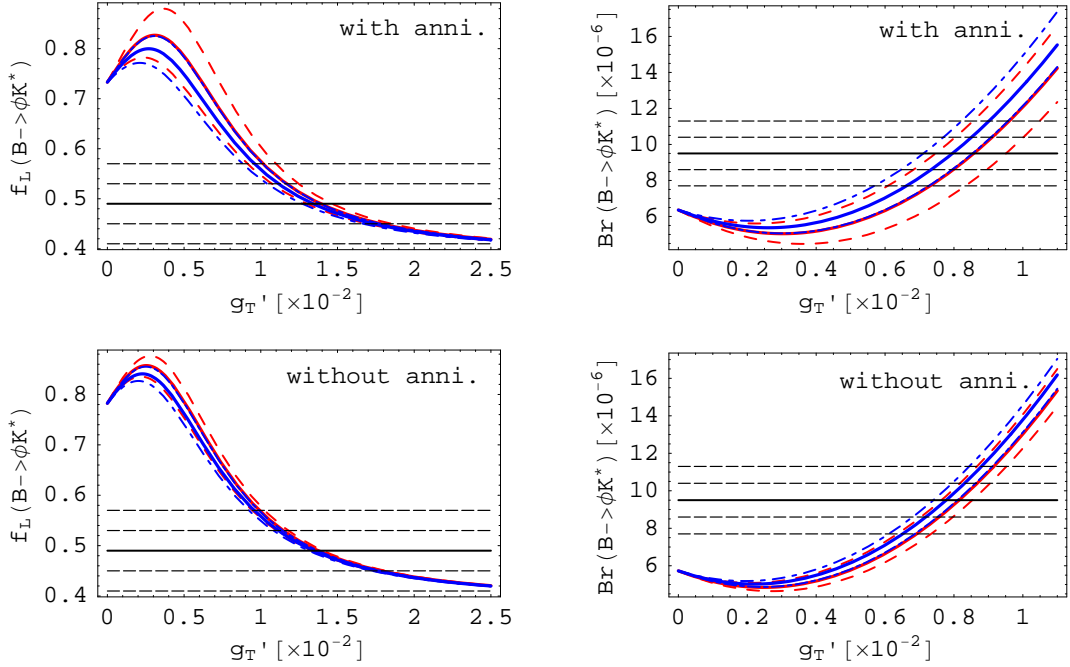}}
\centerline{\parbox{16cm}{\caption{\label{flbr-gt1} \small The
dependence of ${\mathcal B}(B^{0}\to \phi K^{\ast0})$ and
$f_{L}(B^{0}\to \phi K^{\ast0})$ on the parameter $g_{T}^{\prime}$
with the NP weak phase $\delta_{T}$ extracted from the four
$B^{0,+}\to\eta K^{(\ast)0,+}$ decays. The upper and the lower plots
denote the results with and without the annihilation contributions,
respectively. In each plots, the solid blue~(red) curves are the
results with $\delta_{T}$ given by Case I~(II), and dashed curves
due to the error bars of this parameter. The horizontal lines are
the experimental data with the solid lines being the central values
and the dashed ones the error bars~($1\sigma$ and $2\sigma$). }}}
\end{center}
\end{figure}

\begin{figure}[ht]
\begin{center}
\epsfxsize=16cm {\epsffile{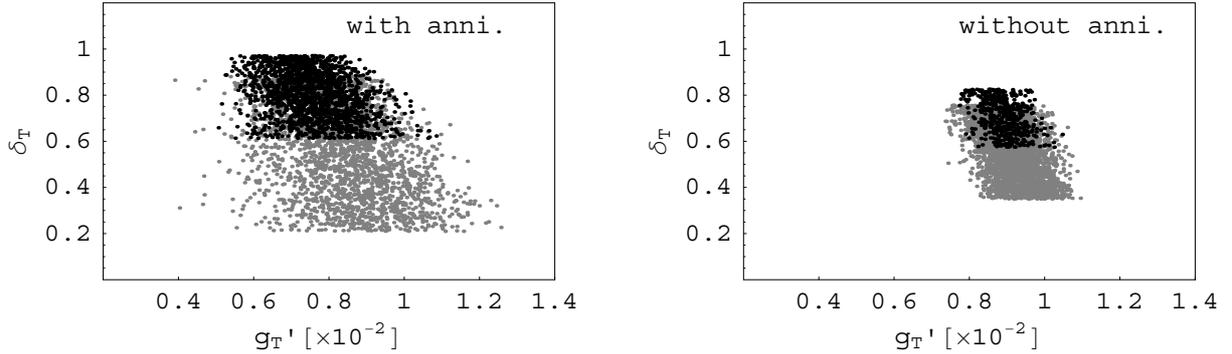}}
\centerline{\parbox{16cm}{\caption{\label{gt1deltat} \small The
allowed regions for the parameters $g_{T}^{\prime}$ constrained by
${\mathcal B}(B^{0}\to\phi K^{\ast0})$ and $f_L$ with the NP weak
phase $\delta_{T}$ extracted from the four $B^{0,+}\rightarrow\eta
K^{(\ast)0,+}$ decays. Other captions are the same as in
Fig.~\ref{gtdeltat2}. }}}
\end{center}
\end{figure}

Now we discuss constraints on the NP parameters from
$B^{0}\rightarrow\phi K^{\ast0}$ decay. With the NP weak phase
$\delta_T$ already extracted from the four $B^{0,+}\rightarrow\eta
K^{(\ast)0,+}$ decays, the parameter $g^{\prime}_{T}$ could be
severely constrained by the well measured ${\mathcal B}(B^{0}\to\phi
K^{\ast0})$ and $f_L$. Our results are presented in
Table~\ref{tab:phik}, Figs.~\ref{flbr-gt1} and \ref{gt1deltat}.

%%%%%%%%%%%%%%%%%%%%%%%%%%%%%%%%%%%%%%%%%
\begin{table}[t]
\centerline{\parbox{16cm}{\caption{\label{tab:phik} \small
Experimental data~\cite{HFAG} and theoretical predictions for the
observables in $B^{0}\to\phi K^{\ast0}$ decay. The other captions
are the same as in Table~\ref{br1-4comparison}. }}}
\begin{center}
\begin{tabular}{lcccc}\hline\hline
Observable & Experiment & SM & Case I & Case II  \\\hline
${\mathcal B}(\times10^{-6})$
  &$9.5\pm0.9$&$5.9^{+1.2}_{-1.0}$&$9.6^{+0.5}_{-0.6}$&$9.9^{+0.8}_{-1.1}$ \\
$$&$$&$5.7\pm0.6$&$10.0\pm0.3$&$10.3\pm0.7$ \\
$f_{L}$
  &$0.49\pm0.04$&$0.76^{+0.06}_{-0.07}$&$0.50^{+0.02}_{-0.02}$&$0.54^{+0.02}_{-0.03}$ \\
$$&$$&$0.78\pm0.03$&$0.52\pm0.01$&$0.55\pm0.01$ \\
$\phi_{\parallel}({\rm rad})$
  &$2.41^{+0.18}_{-0.16}$&$2.90^{+0.25}_{-0.27}$&$1.95^{+0.13}_{-0.12}$&$1.81^{+0.34}_{-0.18}$\\
$$&$$&$2.91\pm0.01$&$1.82\pm0.05$&$1.72\pm0.08$\\
$\phi_{\perp}({\rm rad})$
  &$2.52\pm0.17$&$2.90^{+0.25}_{-0.27}$&$1.96^{+0.13}_{-0.12}$&$1.83^{+0.38}_{-0.18}$\\
$$&$$&$2.91\pm0.01$&$1.82\pm0.05$&$1.73\pm0.08$\\
\hline \hline
\end{tabular}
\end{center}
\end{table}
%%%%%%%%%%%%%%%%%%%%%%%%%%%%%%%%%%%%%%

%%%%%%%%%%%%%%%%%%%%%%%%%%%%%%%%%%%%%%
%%%%%%%%%%%%%%%%%%%%%%%%%%%%%%%%%%%%%%%%

From Table~\ref{tab:phik}, we can see that, with our choice for the
input parameters, especially our quite optimistic choice for the
annihilation contributions, the SM predictions for both ${\mathcal
B}(B^{0}\to\phi K^{\ast0})$ and $f_L$ deviate from the experimental
data, and possible NP scenarios beyond the SM may be needed to
resolve the observed polarization anomaly.

Fig.~\ref{flbr-gt1} shows the dependence of ${\mathcal
B}(B^{0}\to\phi K^{\ast0})$ and $f_L$ on the parameter
$g_{T}^{\prime}$, with the NP weak phase $\delta_{T}$ extracted from
the four $B^{0,+}\to\eta K^{(\ast)0,+}$ decays. To illuminate the
dependence more clearly, we have taken all the other input
parameters to be their center values. As shown in
Fig.~\ref{flbr-gt1}, both ${\mathcal B}(B^{0}\to\phi K^{\ast0})$ and
$f_L$ are very sensitive to the parameter $g_{T}^{\prime}$. It is
particularly interesting to note that the variation trends of these
two observables relative to the parameter $g_{T}^{\prime}$ are
opposite to each other. Thus, with the allowed regions for the NP
weak phase $\delta_{T}$ extracted from the four
$B^{0,+}\rightarrow\eta K^{(\ast)0,+}$ decays, and the experimental
data on these two observables, we could get constraints on the
parameter $g_{T}^{\prime}$. The final allowed regions for the
parameters $g_{T}^{\prime}$ and $\delta_{T}$ are shown in
Fig.~\ref{gt1deltat}, with the corresponding numerical results given
in Table~\ref{gtdeltatresult}.

%%%%%%%%%%%%%%%%%%%%%%%%%%%%%%%%%%%%%%
%%%%%%%%%%%%%%%%%%%%%%%%%%%%%%%%%%%%%%

Corresponding to the two allowed regions for the parameters
$g_{T}^{\prime}$ and $\delta_{T}$ given in
Table~\ref{gtdeltatresult}, both ${\mathcal B}(B^{0}\to\phi
K^{\ast0})$ and $f_L$ are in good agreement with the experimental
data as shown in Table~\ref{tab:phik}. On the other hand, without
the annihilation contributions, our predictions for the relative
phases $\phi_{\parallel}$ and $\phi_{\perp}$ in the decay
$B^{0}\rightarrow\phi K^{\ast0}$ are not consistent with the
experimental data. This mismatch is, however, reduced by including
the annihilation contributions associated with strong phase, which
may indicate the annihilation contributions to be complex. We also
note that the annihilation contributions have been proved recently
to be real and power suppressed~\cite{Arnesen:2006vb}. Moreover, in
Ref.~\cite{Datta}, it is found that the NP strong phases are
generally negligibly small compared to those of the SM
contributions. Further discussion of possible resolution to the
mismatch would be beyond the scope of this paper.

\subsection{Final allowed regions for NP parameters and
predictions for $B_{s}\rightarrow \phi \phi$ decay}

Finally, using the relations
\begin{eqnarray}
g_{T} &=& C_{T8}+\frac{C_{T1}}{N_{c}},\\
g^{\prime}_{T} &=& (1+\frac{1}{2 N_{c}})\,C_{T1}+(
\frac{1}{2}+\frac{1}{N_{c}})\,C_{T8},
\end{eqnarray}
with $N_{c}=3$, we get the final allowed regions for the parameters
$C_{T1}$ and $C_{T8}$, which are presented in
Table~\ref{tab:c15c16result}. It is interesting to note that the
color-octet operator $O_{T8}$ dominates the NP contributions.
Actually, with $O_{T8}$ only, we obtain
$C_{T8}=(7.3^{+1.6}_{-1.5})\times 10^{-3}~((7.7\pm 1.6)\times
10^{-3})$ and $C_{T8}=(9.1^{+1.2}_{-1.0}) \times 10^{-3}~((10.7\pm
0.6) \times 10^{-3})$ from $B\to \eta K^{(\ast)}$ and $B\to\phi
K^{\ast}$ decays, respectively, where (also for the following
results) the numbers in the bracket are obtained without the
annihilation contributions. Thus, we  obtain a single color-octet
$O_{T8}$ solution with $C_{T8}=(8.5^{+0.9}_{-0.9})\times
10^{-3}~((10.3\pm0.5)\times 10^{-3})$ and
$\delta_{T}={44.1^{\circ}}^{+11.6^{\circ}}_{-9.0^{\circ}}~(40.2^{\circ}\pm7.2^{\circ})$.
However, with the color-singlet operator $O_{T1}$ only, we could not
get any solution.

%%%%%%%%%%%%%%%%%%%%%%%%%%%%%%%%%%%%
\begin{table}[t]
\centerline{\parbox{16cm}{\caption{\label{tab:c15c16result} \small
Final results for the coefficients $C_{T1}$, $C_{T8}$, and the weak
phase $\delta_{T}$ extracted from $B\rightarrow\eta K^{(\ast)}$ and
$B^{0}\rightarrow\phi K^{\ast0}$ decays. The other captions are the
same as in Table~\ref{gtdeltatresult}. }}}
\begin{center}
\begin{tabular}{lccccc}\hline\hline
&Allowed region&$C_{T1}(\times10^{-3})$&$C_{T8}(\times10^{-3})$&
$\delta_{T}({\rm rad})$\\\hline
Case I &dark &$1.7^{+1.9}_{-1.7}$&$6.8^{+2.2}_{-1.9}$&$0.77^{+0.20}_{-0.16}$\\
&&$2.8\pm1.6$&$6.7\pm2.1$&$0.70\pm0.13$\\
Case II & gray &$2.6^{+2.8}_{-3.0}$ &$6.4^{+3.3}_{-3.7}$ &$0.60^{+0.29}_{-0.39}$\\
&&$2.8\pm2.2$ &$7.1\pm3.0$ &$0.55\pm0.20$\\
\hline \hline
\end{tabular}
\end{center}
\end{table}
%%%%%%%%%%%%%%%%%%%%%%%%%%%%%%%%%%%%%

It is noted that the two tensor operators $O_{T1}$ and $O_{T8}$
could be expressed by the other two (pseudo-)~scalar operators
$O_{S+P}$ and $O'_{S+P}$ through the Fierz transformation
relations~\cite{KC}
\begin{equation}
O_{S+P}=\frac{1}{12}O_{T1}-\frac{1}{6}O_{T8}, \qquad
O'_{S+P}=\frac{1}{12}O_{T8}-\frac{1}{6}O_{T1}.
\end{equation}
Such operators could be generated in many NP
scenarios with scalar interactions. It would be useful to present
the constraints on the coefficients of the two $(S+P)\otimes(S+P)$
operators. To this end, we get
\begin{eqnarray}
C_{S+P}&=&(-0.99^{+0.39}_{-0.35})\times 10^{-3}~((-0.89\pm
0.38)\times
10^{-3}),\\
C'_{S+P}&=&(0.29^{+0.36}_{-0.32})\times 10^{-3}~((0.09\pm
0.32)\times 10^{-3}),
\end{eqnarray}
with the normalization factor $\frac{G_{F}}{\sqrt 2}|V_{ts}| $ and a
new weak phase
$\delta_{T}={44.1^{\circ}}^{+11.6^{\circ}}_{-9.0^{\circ}}~(40.2^{\circ}\pm7.2^{\circ})$,
corresponding to the two tensor operators case. The single $O_{T8}$
solution would correspond to
$C_{S+P}\equiv-2C^{\prime}_{S+P}=(-1.4^{+0.1}_{-0.2})\times
10^{-3}~((-1.7\pm0.1)\times 10^{-3})$, and the weak phase is the
same as the former. It should be noted that the relation $C_{S+P}=-2
C'_{S+P}$ is strictly required for the last correspondence.

To further test the particular NP scenario with anomalous tensor
operators Eq.~(\ref{eq:np}), we also present our theoretical
predictions for the branching ratio and the longitudinal polarization
fraction of $B_{s}\rightarrow \phi \phi$ decay, as summarized in
Table~\ref{tab:phiphi}. Within the SM, its decay amplitude is given
by~\cite{Beneke:2006hg}
\begin{eqnarray}\label{Bsphiphi-SM}
\frac{1}{2} A_{\overline{B}_s\to\phi\phi}^{\rm SM} &=& A_{\phi\phi}
\sum_{p=u,c}V_{pb}V_{ps}^{\ast} \left[ \alpha_3^{p,h} +
\alpha_4^{p,h} - \frac{1}{2} \alpha_{3,{\rm EW}}^{p,h}- \frac{1}{2}
\alpha_{4,{\rm EW}}^{p,h} + \beta_3^{p,h} - \frac{1}{2}
\beta_{3,{\rm EW}}^{p,h}\right. \nonumber\\
&&\qquad \qquad \mbox{} \left. + \beta_4^{p,h} - \frac{1}{2}
\beta_{4,{\rm EW}}^{p,h}\right].
\end{eqnarray}
With respect to the relevant quantities in Eq.~(\ref{Bsphiphi-SM}),
one can get them directly from those for $B^{0}\rightarrow\phi
K^{\ast0}$ decay with some simple changes. This decay mode is of
particular interest to test the proposed resolutions to the
polarization anomalies observed in $B\to\phi K^{\ast}$ decays, since
both of these decay modes are mediated by the same quark level
subprocess $b\to s\bar{s}s$. In addition, since the two final-state
mesons are identical in this decay mode, more observables in the
time-dependent angular analysis will become
zero~\cite{transversebasis}. So, this decay mode can be considered
as an ideal probe for various NP scenarios proposed to resolve the
polarization anomalies observed in $B\to\phi K^{\ast}$ decays. The
earlier studies of this particular decay mode within the QCDF
formalism have been carried out in Ref.~\cite{Li:2003he}, but
without taking into account the annihilation contributions.

%%%%%%%%%%%%%%%%%%%%%%%%%%%%%%%%%%%%%%%%%%%
\begin{table}[t]
\centerline{\parbox{16cm}{\caption{\label{tab:phiphi} \small
Theoretical predictions for $B_{s}\rightarrow\phi\phi$ decay both
within the SM and in the particular NP scenario Eq.~(\ref{eq:np}).
The other captions are the same as in Table~\ref{br1-4comparison}.
}}}
\begin{center}
\begin{tabular}{lcccc}\hline\hline
Observable & Experiment & SM & Case I & Case II  \\\hline
${\mathcal B}(\times10^{-6})$&$14^{+8}_{-7}$&$20.6^{+4.2}_{-3.0}$&
$29.4^{+4.4}_{-3.5}$&$20.1^{+5.9}_{-4.8}$\\
$$&$$&$17\pm2$&$30\pm3$&$29\pm3$\\
$f_{L}$&---&$0.83^{+0.04}_{-0.06}$&$0.73^{+0.07}_{-0.08}$&
$0.72^{+0.07}_{-0.07}$\\
$$&&$0.83\pm0.02$&$0.66\pm0.03$&$0.67\pm0.03$ \\
\hline \hline
\end{tabular}
\end{center}
\end{table}
%%%%%%%%%%%%%%%%%%%%%%%%%%%%%%%%%%%%%%%%%%%%%%

From the numerical results presented in Tables~\ref{tab:phik} and
\ref{tab:phiphi}, we can see that both the branching ratio and the
longitudinal polarization fraction of $B_{s}\rightarrow \phi \phi$
decay is larger than those of the decay $B^{0}\rightarrow\phi
K^{\ast0}$, which is due to the relative factor of two in the $\phi
\phi$ amplitude Eq.~(\ref{Bsphiphi-SM}), as well as the additional
contribution from the annihilation coefficient $\beta_4^{p,h}$. In
particular, due to an accidental cancelation, the annihilation
coefficient $\beta_3^{p,0}$~(contributing to both
$H_{00}(B^{0}\to\phi K^{\ast 0})$ and $H_{00}(B_{s}\to\phi\phi)$) is
quite smaller than $\beta_4^{p,0}$~\cite{Beneke:2006hg}.
Interestingly, recent calculations made in the perturbative QCD
approach also predict a large branching ratio ${\mathcal
B}(B_{s}\rightarrow \phi
\phi)=(44.1^{+8.3+13.3+0.0}_{-6.5-~8.4-0.0})\times 10^{-6}$ with the
longitudinal polarization fraction $f_L(B_{s}\rightarrow
\phi\phi)=(68.0^{+4.2+1.7+0.0}_{-4.0-2.2-0.0})\times
10^{-2}$~\cite{C.D}. Their result for $f_L(B_{s}\rightarrow
\phi\phi)$ is smaller than our predictions within the SM. The
predicted results presented in Table~\ref{tab:phiphi} could be
tested at the Fermilab Tevatron and the LHC-b experiments.

\section{Conclusions}

In summary, motivated by the observed discrepancies between the
experimental data and the SM predictions for the branching ratios of
$B\rightarrow \eta K^{\ast}$ decays and the polarization fractions
in $B\rightarrow \phi K^{\ast}$ decays, we have studied a particular
NP scenario with anomalous tensor operators $O_{T1}$ and $O_{T8}$,
which has been proposed to resolve the polarization anomalies in the
literature~\cite{kagan,KC,YY}.

After extensive numerical analysis, we have found that the above
observed discrepancies could be resolved simultaneously and
constraints on the NP parameters have been obtained. With both the
experimental data and the theoretical input parameters varying
within $1\sigma$ error bars, we have found the following two
solutions: (I) both the two tensor operators contribute with the
parameter spaces presented in the upper two rows of
Table~\ref{tab:c15c16result}; (II) only $O_{T8}$ contributes with
$C_{T8}=(8.5^{+0.9}_{-0.9})\times 10^{-3}((10.3\pm0.5)\times
10^{-3})$ and $\delta_{T}$ the same as solution (I). The above two
solutions correspond to the scenario with new operators $O_{S+P}$
and $O^{\prime}_{S+P}$ added to the SM, with the parameter spaces:
(I) $C_{S+P}=(-0.99^{+0.39}_{-0.35})\times
10^{-3}~((-0.89\pm0.38)\times 10^{-3})$,
$C^{\prime}_{S+P}=(0.29^{+0.36}_{-0.32})\times
10^{-3}~((0.09\pm0.32)\times 10^{-3})$ and (II)
$C_{S+P}\equiv-2C^{\prime}_{S+P}=(-1.4^{+0.1}_{-0.2})\times
10^{-3}~((-1.7\pm0.1)\times 10^{-3})$. Their weak phase
$\delta_{S+P}$ is the same as $\delta_{T}$.

Against our early expectations, we have found that the solution with
two tensor operators are dominated by the color-octet operator
$O_{T8}$. It would be interesting to investigate whether the
available NP models could give such effective interactions at the
$m_{b}$ scale.

To further test the particular NP scenario with two anomalous tensor
operators $O_{T1,T8}$ or (pseudo-)~scalar operators
$O^{(\prime)}_{S+P}$, we have also presented our predictions for the
observables in $B_{s}\rightarrow \phi \phi$ decay, which could be
tested more thoroughly at the Fermilab Tevatron and the LHC-b
experiments in the near future.

It should be noted that the strong constraints in
Table~\ref{tab:c15c16result} obtained without annihilation
contributions may be too optimistic. As has been shown in the table,
the uncertainties due to poorly known annihilation contributions
could dilute the requirement of NP very much. Generally, this caveat
could be applied to probe possible NP scenarios in exclusive
non-leptonic $B$ decays. Further theoretical progress is strongly
demanded.

\section*{Acknowledgments}

The work is supported by National Science Foundation under contract
Nos.~10305003 and 10675039, and the NCET Program sponsored by
Ministry of Education, China.

\begin{appendix}

\section*{Appendix~A: Decay amplitudes in the SM with QCDF}

The amplitudes for $B\to \eta K^{(\ast)}$ are recapitulated from
Ref.~\cite{beneke3}
\begin{eqnarray}
\sqrt2\,{\cal A}_{B^-\to K^-\eta}^{\rm SM}
   &=& \sum_{p=u,c}V_{pb}V_{ps}^{\ast} \biggl\{ A_{K^-\eta_q} \Big[
    \delta_{pu}\,\alpha_2 + 2\alpha_3^p + \half\alpha_{3,{\rm
    EW}}^p\Big]\biggl. \nonumber\\
  && \biggl. +\sqrt{2} A_{K^-\eta_s} \Big[\delta_{pu}\,\beta_2 + \alpha_3^p
    + \alpha_4^p - \half\alpha_{3,{\rm EW}}^p -\half\alpha_{4,{\rm EW}}^p
    + \beta_3^p +\beta_{3,{\rm EW}}^p\Big]\biggl. \nonumber\\
  && \biggl. + A_{\eta_q K^-} \Big[ \delta_{pu}\,(\alpha_1+\beta_2)
    + \alpha_4^p+\alpha_{4,{\rm EW}}^p
    + \beta_3^p +\beta_{3,{\rm
    EW}}^p\Big]\biggl\}\,,\label{amp1_SM}
\end{eqnarray}
\begin{eqnarray}
\sqrt2\,{\cal A}_{\overline B^0\to\overline K^0\eta}^{\rm SM}
   &=& \sum_{p=u,c}V_{pb}V_{ps}^{\ast} \biggl\{ A_{\bar K^0\eta_q}
   \Big[ \delta_{pu}\,\alpha_2 + 2\alpha_3^p + \half\alpha_{3,{\rm EW}}^p\Big]
   \biggl. \nonumber\\
   && \biggl. +\sqrt{2} A_{\bar K^0\eta_s} \Big[ \alpha_3^p + \alpha_4^p
    - \half\alpha_{3,{\rm EW}}^p - \half\alpha_{4,{\rm EW}}^p + \beta_3^p
    - \half\beta_{3,{\rm EW}}^p \Big]\biggl. \nonumber\\
    && \biggl. + A_{\eta_q\bar K^0} \Big[\alpha_4^p - \half\alpha_{4,{\rm EW}}^p
    + \beta_3^p - \half\beta_{3,{\rm EW}}^p\Big] \biggl\}\,,\label{amp2_SM}
\end{eqnarray}
\begin{eqnarray}
\sqrt2\,{\cal A}_{B^-\to K^{\ast-}\eta}^{\rm SM}
   &=& \sum_{p=u,c}V_{pb}V_{ps}^{\ast} \biggl\{ A_{K^{\ast-}\eta_q} \Big[
    \delta_{pu}\,\alpha_2 + 2\alpha_3^p + \half\alpha_{3,{\rm
    EW}}^p\Big]\biggl. \nonumber\\
  && \biggl. +\sqrt{2} A_{K^{\ast-}\eta_s} \Big[\delta_{pu}\,\beta_2
    + \alpha_3^p + \alpha_4^p - \half\alpha_{3,{\rm EW}}^p
    -\half\alpha_{4,{\rm EW}}^p +\beta_3^p +\beta_{3,{\rm
    EW}}^p\Big]\biggl. \nonumber\\
  && \biggl. +A_{\eta_q K^{\ast-}} \Big[ \delta_{pu}\,(\alpha_1
    +\beta_2)+ \alpha_4^p+\alpha_{4,{\rm EW}}^p
    + \beta_3^p +\beta_{3,{\rm EW}}^p\Big]\biggl\}\,,\label{amp3_SM}
\end{eqnarray}
\begin{eqnarray}
\sqrt2\,{\cal A}_{\overline{B}^0\to\overline{K}^{\ast0}\eta}^{\rm
SM}
   &=& \sum_{p=u,c}V_{pb}V_{ps}^{\ast} \biggl\{ A_{\bar K^{\ast0}\eta_q}
   \Big[\delta_{pu}\,\alpha_2 + 2\alpha_3^p + \half\alpha_{3,{\rm EW}}^p\Big]
   \biggl. \nonumber\\
   && \biggl. +\sqrt{2} A_{\bar K^{\ast0}\eta_s} \Big[\alpha_3^p
    + \alpha_4^p - \half\alpha_{3,{\rm EW}}^p
    -\half\alpha_{4,{\rm EW}}^p + \beta_3^p -
    \half\beta_{3,{\rm EW}}^p\Big]\biggl. \nonumber \\
   && \biggl. + A_{\eta_q\bar K^{\ast0}} \Big[\alpha_4^p -\half\alpha_{4,{\rm EW}}^p
    + \beta_3^p -\half\beta_{3,{\rm EW}}^p\Big] \biggl\}\,,\label{amp4_SM}
\end{eqnarray}
where the explicit expressions for the coefficients
$\alpha_i^p\equiv\alpha_i^p(M_1M_2)$ and
$\beta_i^p\equiv\beta_i^p(M_1M_2)$ could also be found in
Ref.~\cite{beneke3}.  We recall that the $\alpha_{i}^p$ terms
contain one-loop vertex, penguin and hard spectator contributions,
whereas the $\beta_{i}^p$ terms are due to the weak annihilation,
and the transition form factors are encoded in the factors
$A_{M_{1}M_{2}}$.

The decay amplitude of $\overline{B}^{0}\rightarrow \phi
\overline{K}^{\ast0}$ mode can be read off from
Refs.~\cite{Beneke:2006hg,BenVV}
\begin{equation}\label{eq:phiksm}
{\cal A}_{\overline{B}^{0}\rightarrow\phi \overline{K}^{\ast0}}^{\rm
SM} = A_{\bar{K}^{\ast0}\phi}^h \sum_{p=u,c}V_{pb}V_{ps}^{\ast}\Big[
    \alpha_3^{p,h} + \alpha_4^{p,h}- \half\alpha_{3,{\rm EW}}^{p,h}
    -\half\alpha_{4,{\rm EW}}^{p,h} + \beta_3^{p,h} - \half\beta_{3,{\rm
    EW}}^{p,h}\Big]\,,
\end{equation}
where the superscript `$h$' denotes the helicity of two final-state
vector mesons, with $h=0,+,-$ corresponding to two outgoing
longitudinal, positive, and negative helicity vector mesons,
respectively.

In the helicity basis, the decay amplitude Eq.~(\ref{eq:phiksm}) can
be further decomposed into the following three helicity amplitudes
\begin{eqnarray}
H_{00}^{\rm SM}&=&A_{\bar{K}^{\ast0}\phi}^0
    \sum_{p=u,c}V_{pb}V_{ps}^{\ast} \Big[
    \alpha_3^{p,0} + \alpha_4^{p,0}- \half\alpha_{3,{\rm EW}}^{p,0}
    -\half\alpha_{4,{\rm EW}}^{p,0} + \beta_3^{p,0} - \half\beta_{3,{\rm
    EW}}^{p,0}\Big],\label{hphi1_SM}\\
H_{\pm\pm}^{\rm SM}&=&A_{\bar{K}^{\ast0}\phi}^{\pm}
    \sum_{p=u,c}V_{pb}V_{ps}^{\ast} \Big[
    \alpha_3^{p,{\pm}} + \alpha_4^{p,{\pm}}- \half\alpha_{3,{\rm EW}}^{p,{\pm}}
    -\half\alpha_{4,{\rm EW}}^{p,{\pm}} + \beta_3^{p,{\pm}} - \half\beta_{3,{\rm
    EW}}^{p,{\pm}}\Big],\label{hphi2_SM}
\end{eqnarray}
with~\cite{Beneke:2006hg}
\begin{equation}
A_{\bar{K}^{\ast0}\phi}^h \equiv \frac{G_F}{\sqrt{2}}\,
        \langle \overline{K}^{\ast0}|(\bar{s} b)_{V-A}|\overline{B}^{0} \rangle
        \langle \phi |(\bar{s} s)_{V}|0\rangle\,,
\end{equation}
\begin{eqnarray} \label{eq:Aexpl}
A^0_{\bar{K}^{\ast0}\phi} &=& \frac{i G_F}{\sqrt{2}} \frac{m_{B_d}^3
f_{\phi}}{2m_{K^{\ast 0}}}\, \left[(1+\frac{m_{K^{\ast
0}}}{m_{B_d}}) A_1^{B\to K^{\ast}}(m_{\phi}^2) - (1-\frac{m_{K^{\ast
0}}}{m_{B_d}}) A_2^{B\to K^{\ast}}(m_{\phi}^2)\right] \,,\\
A^{\pm}_{\bar{K}^{\ast 0}\phi} &=& \frac{i G_F}{\sqrt{2}} m_{B_d}
m_{\phi} f_{\phi}\, \left[(1+\frac{m_{K^{\ast 0}}}{m_{B_d}})
A_1^{B\to K^{\ast}}(m_{\phi}^2)\mp (1-\frac{m_{K^{\ast
0}}}{m_{B_d}}) V^{B\to K^{\ast}}(m_{\phi}^2)\right].
\end{eqnarray}

\section*{Appendix~B: Theoretical input parameters}

\subsection*{B1. Wilson coefficients and CKM matrix elements}

The Wilson coefficients $C_{i}(\mu)$ have been evaluated reliably to
next-to-leading logarithmic order~\cite{Buchalla:1996vs,Buras:2000}.
Their numerical results in the naive dimensional regularization
scheme at the scale $\mu=m_{b}$~($\mu_h=\sqrt{\Lambda_h m_b}$) are
given by
\begin{eqnarray}
&&C_{1}=1.077~(1.178), \quad C_{2}=-0.174~(-0.355), \quad
  C_{3}=0.013~(0.027), \nonumber \\
&&C_{4}=-0.034~(-0.060), \quad C_{5}=0.008~(0.011), \quad
  C_{6}=-0.039~(-0.081),\nonumber \\
&&C_{7}/\alpha_{e.m.}=-0.013~(-0.034), \quad C_{8}/\alpha_{e.m.}=0.047~(0.099), \quad
  C_{9}/\alpha_{e.m.}=-1.208~(-1.338), \nonumber \\
&&C_{10}/\alpha_{e.m.}=0.229~(0.426), \quad
C_{7\gamma}=-0.297~(-0.360), \quad C_{8g}=-0.143~(-0.168).
\end{eqnarray}
The values at the scale $\mu_{h}$, with $m_{b}=4.79~{\rm GeV}$ and
$\Lambda_{h}=500~{\rm MeV}$, should be used in the calculation of
hard-spectator and weak annihilation contributions.

For the CKM matrix elements, we adopt the Wolfenstein
parameterization~\cite{Wolfenstein:1983yz} and choose the four
parameters $A$, $\lambda$, $\rho$, and $\eta$
as~\cite{Charles:2004jd}
\begin{equation}
A=0.809\pm 0.014, \quad \lambda=0.2272\pm 0.0010, \quad
\overline{\rho}=0.197^{+0.026}_{-0.030}, \quad
\overline{\eta}=0.339^{+0.019}_{-0.018},
\end{equation}
with $\overline{\rho}=\rho\,(1-\frac{\lambda^2}{2})$ and
$\bar{\eta}=\eta\,(1-\frac{\lambda^2}{2})$.

\subsection*{B2. Quark masses and lifetimes}

As for the quark mass, there are two different classes appearing in
our calculation. One type is the pole quark mass appearing in the
evaluation of penguin loop corrections, and denoted by $m_q$. In
this paper, we take
\begin{equation}
 m_u=m_d=m_s=0, \quad m_c=1.64\,{\rm GeV}, \quad m_b=4.79\,{\rm GeV}.
\end{equation}
The other one is the current quark mass which appears in the factor
$r_\chi^M$ through the equation of motion for quarks. This type of
quark mass is scale dependent and denoted by $\overline{m}_q$. Here
we take~\cite{PDG06}
\begin{equation}
\overline{m}_s(\mu)/\overline{m}_q(\mu)= 25\sim30\,,\quad
\overline{m}_{s}(2\,{\rm GeV}) = (95\pm25)\,{\rm MeV}\,, \quad
\overline{m}_{b}(\overline{m}_{b})= 4.20\,{\rm GeV}\,,
\end{equation}
where $\overline{m}_q(\mu)=(\overline{m}_u+\overline{m}_d)(\mu)/2$,
and the difference between $u$ and $d$ quark is not distinguished.

As for the lifetimes of $B$ mesons, we take~\cite{PDG06}
$\tau_{B_{u}} = 1.638\,{\rm ps}$, $ \tau_{B_{d}}=1.530\,{\rm ps}$,
and $ \tau_{B_{s}}=1.466\,{\rm ps}$ as our default input values.

\subsection*{B3. The decay constants and form factors}
 In this paper, we take the decay constants
\begin{eqnarray}
 & &f_{B}=(216\pm22)~{\rm MeV}~\cite{Gray:2005ad}, \quad
    f_{B_s}=(259\pm32)~{\rm MeV}~\cite{Gray:2005ad}, \quad
    f_\pi=(130.7\pm0.4)~{\rm MeV}~\cite{PDG06},\nonumber\\
 & &f_{K}=(159.8\pm1.5)~{\rm MeV}~\cite{PDG06} \quad
    f_{K^{\ast}}=(217\pm5)~{\rm MeV}~\cite{BallZwicky}, \quad
    f_{\phi}=(231\pm4)~{\rm MeV}~\cite{BallZwicky}, \nonumber\\
 & &f_{K^{\ast}}^{\perp}(1~{\rm GeV}) =(185\pm10)~{\rm MeV}~\cite{BallZwicky}, \quad
    f_{\phi}^{\perp}(1~{\rm GeV}) =(200\pm10)~{\rm MeV}~\cite{BallZwicky},
\end{eqnarray}
and the form factors~\cite{BallZwicky}
\begin{eqnarray}
 & &F^{B\to \eta}_{0}(0)=0.275\pm0.036, \quad
    F^{B\to {K}}_{0}(0)=0.331\pm0.041, \quad
    A_0^{B\to K^\ast}(0)=0.374\pm0.033, \nonumber\\
 & &V^{B\to K^\ast}(0)=0.411\pm0.033, \quad
    A_1^{B\to K^\ast}(0)=0.292\pm0.028, \quad
    A_2^{B\to K^\ast}(0)=0.259\pm0.027, \nonumber\\
 & &V^{B_s\to \phi}(0)=0.434\pm0.035, \quad
    A_1^{B_s\to \phi}(0)=0.311\pm0.030, \quad
    A_2^{B_s\to \phi}(0)=0.234\pm0.028, \nonumber\\
 & &T_1^{B\to K^\ast}(0)=0.333\pm0.028, \quad
    T_2^{B\to K^\ast}(0)=0.333\pm0.028, \quad
    T_1^{B_s\to \phi}(0)=0.349\pm0.033, \nonumber\\
 & &T_2^{B_s\to \phi}(0)=0.349\pm0.033.
\end{eqnarray}
For the parameters related to the $\eta$-$\eta^{\prime}$ mixing, we
choose~\cite{FKS}
\begin{equation}
f_q=(1.07\pm0.02)\, f_\pi, \quad f_s=(1.34\pm0.06)\,f_\pi, \quad
\phi=39.3^{\circ}\pm1.0^{\circ}\,.
\end{equation}
The other parameters relevant to the meson $\eta$ can be obtained
from the above three ones~\cite{beneke1}.

\subsection*{B4. The LCDAs of mesons.}

For the LCDAs of mesons, we use their asymptotic
forms~\cite{projector,formfactor}
\begin{eqnarray}
&&\Phi_P(x)=\Phi_{\parallel,\perp}^V(x)=g_{\perp}^{(a)V}(x)=6\,x(1-x)\,,
\quad
\Phi_p (x)=1,\nonumber\\
&&\Phi_v (x) =3\,(2\,x-1)\,,\qquad
g_{\perp}^{(v)V}(x)=\frac{3}{4}\left[1+(2\,x-1)^2\,\right].
\end{eqnarray}

As for the $B$ meson wave functions, we need only consider the first
inverse moment of the leading LCDA $\Phi_1^B (\xi)$ defined
by~\cite{beneke2}
\begin{equation}\label{PhiB1}
  \int_0^1 \frac{d\xi}{\xi}\,\Phi_1^B(\xi)
  \equiv \frac{m_B}{\lambda_B}\,,
\end{equation}
where $\lambda_B=(460\pm110)\,{\rm MeV}$~\cite{Braun} is the
hadronic parameter introduced to parameterize this integral.

\end{appendix}

%%%%%%%%%%%%%%%%%%%%%%%%%%%%%%

\end{document}